\documentclass[preprint,flushrt]{aastex}

\newcommand{\reffig}[1]{Figure \ref{#1}}
\newcommand{\reftbl}[1]{Table \ref{#1}}

\slugcomment{To appear in \apj}
\shorttitle{PHOTOMETRIC REDSHIFTS OF HDF--S NICMOS}
\shortauthors{Yahata et al.}

\begin{document}

\title{Photometry and Photometric Redshifts of Faint Galaxies in the Hubble
Deep Field South NICMOS Field$^{1,2}$}

\author{Noriaki Yahata, Kenneth M. Lanzetta, Hsiao-Wen Chen,\\
Alberto Fern\'andez-Soto, Sebastian M. Pascarelle, and Amos Yahil}
\affil{Department of Physics and Astronomy, State University of New York
at Stony Brook,\\ Stony Brook, NY 11794--3800, U.S.A.}
\and
\author{Richard C. Puetter}
\affil{Center for Astrophysics and Space Sciences, University of California
at San Diego,\\ La Jolla, CA 92093--0424, U.S.A.}

\altaffiltext{1}
{Based on observations with the NASA/ESA Hubble Space Telescope, obtained at
the Space Telescope Science Institute, which is operated by the Association of
Universities for Research in Astronomy, Inc., under NASA contract NAS 5--26555.}

\altaffiltext{2}
{Based on observations collected at the European Southern Observatory, Paranal,
Chile (VLT-UT1 Science Verification Program).}


\begin{abstract}
We present a catalog of photometry and photometric redshifts of 335 faint
objects in the HDF--S NICMOS field.  The analysis is based on (1) infrared
images obtained with the Hubble Space Telescope (HST) using the Near Infrared
Camera and Multi-Object Spectrograph (NICMOS) with the F110W, F160W, and F222M
filters, (2) an optical image obtained with HST using the Space Telescope Imaging
Spectrograph (STIS) with no filter, and (3) optical images obtained with the
European Southern Observatory (ESO) Very Large Telescope (VLT) with {\it U},
{\it B}, {\it V}, {\it R}, and {\it I\/} filters.  The primary utility of the
catalog of photometric redshifts is as a survey of faint galaxies detected in
the NICMOS F160W and F222M images.  The sensitivity of the survey varies
significantly with position, reaching a limiting depth of $AB(16,000) \approx
28.7$ and covering 1.01 arcmin$^2$ to $AB(16,000) = 27$ and 1.05 arcmin$^2$ to
$AB(16,000) = 26.5$.  The catalog of photometric redshifts identifies 21 galaxies
(or 6\% of the total) of redshift $z > 5$, 8 galaxies (or 2\% of the total) of
redshift $z > 10$, and 11 galaxies (or 3\% of the total) of best-fit spectral
type E/S0, of which 5 galaxies (or 1\% of the total) are of redshift $z > 1$.
\end{abstract}


\clearpage
\section{INTRODUCTION}

  The Hubble Deep Field South (HDF--S) images are among the deepest images of
the universe ever obtained at optical and infrared wavelengths.  In this paper,
we present a catalog of photometry and photometric redshifts of 335 faint
objects in the HDF--S NICMOS field.  The analysis is based on (1) infrared
images obtained with the Hubble Space Telescope (HST) using the Near Infrared
Camera and Multi-Object Spectrograph (NICMOS) with the F110W, F160W, and F222M
filters, (2) an optical image obtained with HST using the Space Telescope
Imaging Spectrograph (STIS) with no filter, and (3) optical images obtained
with the European Southern Observatory (ESO) Very Large Telescope (VLT) with
{\it U}, {\it B}, {\it V}, {\it R}, and {\it I\/} filters.  The analysis is similar
to the analyses of the Hubble Deep Field (HDF) described previously by
\citet[hereafter LYF96]{LYF96} and \citet[hereafter FLY99]{FLY99}, although
in detail the current analysis differs from the previous analyses in three
important ways:

  First, objects are detected in the NICMOS F160W and F222M images, at central
wavelengths of $\lambda \approx 16,000$ \AA\ and $\lambda \approx 22,200$ \AA,
respectively.  The analysis is in principle sensitive to galaxies of redshift
as large as $z \approx 18$, beyond which the Ly$\alpha$-forest absorption
discontinuity is redshifted past the response of the NICMOS F222M filter.

  Second, the optical and infrared photometry is measured using a new
quasi-optimal photometry technique that fits model spatial profiles of detected
objects to the space- and ground-based images.  The technique is based on but
extends the spatial profile fitting technique described previously by FLY99.
In comparison with conventional methods, the new technique provides higher
signal-to-noise ratio measurements, and in contrast with conventional methods,
the new technique accounts for uncertainty correlations between nearby,
overlapping neighbors.

  Third, the photometric redshifts are measured using our redshift likelihood
technique with a sequence of six spectrophotometric templates, including the
four templates of our previous analyses (of E/S0, Sbc, Scd, and Irr galaxies)
and two new templates (of star-forming galaxies).  Inclusion of the two new
templates eliminates the tendency of our previous analyses to systematically
underestimate the redshifts of galaxies of redshift $2 < z < 3$ (by a redshift
offset of roughly 0.3), in agreement with results found previously by 
\citet{Benitez98}.  Comparison with spectroscopic redshifts of galaxies identified
in the HDF and HDF--S indicates that with the sequence of six spectrophotometric
templates the photometric redshifts are accurate to within an RMS relative
uncertainty of $\Delta z/(1 + z) \lesssim 7\%$ at all redshifts $z < 6$ that
have as yet been examined. 

  The primary utility of the catalog of photometric redshifts is as a survey of
faint galaxies detected in the NICMOS F160W and F222M images.  The sensitivity of
the survey varies significantly with position, reaching a limiting depth of
$AB(16,000) \approx 28.7$ and covering 1.01 arcmin$^2$ to $AB(16,000) = 27$ and
1.05 arcmin$^2$ to $AB(16,000) = 26.5$.  Likewise, the survey reaches a limiting
depth of $AB(22,200) \approx 24.8$ and covering 0.79 arcmin$^2$ to
$AB(22,200) = 24$ and 1.09 arcmin$^2$ to $AB(22,200) = 23$.

  The organization of the paper is as follows:  In \S\ 2, the observations are
described.  In \S\ 3, the object identification, photometry, and photometric
redshift measurements are described.  The results are presented in \S\ 4, and
the discussion is presented in \S\ 5 and \S\ 6.  The summary and conclusions
are given in \S\ 7.  Scientific analysis of the catalog will be presented in
forthcoming papers.

\section{OBSERVATIONS}

  The HDF--S NICMOS field is centered at J2000 coordinates $\alpha =$
22:\-32:\-51.75 and $\delta =$ $-$60:\-38:\-48.20.  The observations consist of
three sets of images:  (1) infrared images obtained with HST using NICMOS, (2)
an optical image obtained with HST using STIS, and (3) optical images obtained
with the ESO VLT using the Test Camera.  \reftbl{tbl:obs} summarizes details of
the observations.

  The HST NICMOS images were acquired in September, 1998 using NICMOS with
Camera 3 and the F110W, F160W, and F222M filters.  For each band, the
observations consisted of $\approx 100$ dithered exposures of between 512 s and
1472 s duration.  The raw images were processed and reduced by the Space
Telescope Science Institute (STScI) NICMOS team.  The processed images were
registered onto a grid of $1100 \times 1300$ pixel$^2$ at a scale of 0.075 arcsec
pixel$^{-1}$, which covers an angular area of $\approx 1 \times 1$ arcmin$^2$.
The HST STIS image was acquired in September and October 1998 using STIS with
the 50CCD in open filter mode (which is sensitive at wavelengths spanning
$\lambda \approx 2000 - 10000$ \AA).  The observations consisted of 9 dithered
exposures of 2900 s duration.  The raw images were processed and reduced by the
STScI STIS team.  The processed image was registered onto a grid of $3300
\times 3900$ pixel$^2$ at a scale of 0.025 arcsec pixel$^{-1}$, which covers
roughly the same angular area as the NICMOS images.  We used the non-drizzled
Version 1 release of the NICMOS and STIS images, which were made available by
STScI on 23 November, 1998, and we adopted photometric zero points determined
by the STScI NICMOS and STIS teams.

  The ESO VLT images were acquired in August, 1998 using the Unit Telescope \#1
(UT1) with the Test Camera and the {\it U}, {\it B}, {\it V}, {\it R}, and
{\it I\/} filters as a part of the VLT science verification campaign (ESO VLT-UT1
Science Verification 1998).  For each band, the observations consisted of
$\approx 20$ dithered exposures of $\approx 900$ s duration.  We reduced the raw
images, taking extra care in constructing the flat-field images, because the Test
Camera CCD suffers from a large, wavelength-dependent blemish in its center.
In agreement with \citet{Fontana99}, we found that separate superflats constructed
from the median of the images obtained each night work best.  The processed images
are sampled on a grid of $\approx 1000 \times 1000$ pixel$^2$ at a scale of 0.091
arcsec pixel$^{-1}$, which covers an angular area of $\approx 1.5 \times 1.5$
arcmin$^2$.  We adopted photometric zero points determined by \citet{Fontana99}.

  The point spread functions (PSFs) of the ground-based images vary
significantly from image to image, with the best images characterized by ${\rm
FWHM} \approx 0.5$ arcsec and the worst images characterized by ${\rm FWHM}
\approx 2.2$ arcsec.  For this reason, we worked only with the individual
images, i.e.\ without combining the images in each band.

  We registered the ground-based images to the space-based images (because the
space-based images were already registered to within $\approx 0.05$ pixel by
the STScI NICMOS and STIS teams).  We registered the images by measuring
coordinates of stars in the space- and ground-based images, using these
measurements to derive transformations to the space-based frame, and shifting,
rotating, and scaling the ground-based images to the space-based frame.
Special treatment was required for the {\it U\/}-band images, for which only
a single bright star is available.  In this case, we shifted according to
measurements of the star and rotated and scaled according to measurements from
the other bands.  Adjacent pixels of the final, registered images are correlated
as a result of the reduction and registration procedures, which must be accounted
for by the analysis.

\section{ANALYSIS}

\subsection{Object Detection}

  We detected objects in the NICMOS F160W image following procedures
similar to those described previously by LYF96 and FLY99.  First, we formed a
signal-to-noise image by dividing the F160W image by the square root of the
F160W variance image.  Next, we applied the SExtractor object detection program
{\citep{Bertin96}} to detect objects in the signal-to-noise image.  We set the
SExtractor detection parameters by requiring that no spurious objects were
detected in the ``negative'' signal-to-noise image, which we formed by dividing
the negative of the F160W image by the square root of the F160W variance image.
We detected objects according to a signal-to-noise criterion (rather than a signal
criterion) because the sensitivity of the F160W image varies significantly with
position.  (Generally, the image is more sensitive toward the center and less
sensitive toward the edges as a consequence of the way the individual exposures
were dithered.)  Finally, we modified the resulting SExtractor segmentation map
to make three small corrections:  (1) we eliminated objects near the edges of
the image, (2) we deblended objects around bright stars or galaxies, and (3) we
merged diffraction spikes of bright stars to the host stars.  A total of 332
objects were detected, the brightest of which is of magnitude
$AB(16,000) \approx 17.2$ and the faintest of which is of magnitude
$AB(16,000) \approx 29.2$.

We repeated the same detection procedures using the NICMOS F222M image.  Three
additional objects were detected and incorporated into the object catalog.

\subsection{Photometry}

  We measured optical and infrared photometry using a new quasi-optimal
photometry technique that fits model spatial profiles of the detected objects
to the space- and ground-based images.  The technique is based on but extends
the spatial profile fitting technique described previously by FLY99.  In
particular, the current technique implements two improvements over the previous
technique:  First, we used an image reconstruction method to generate smooth
models of the intrinsic spatial profiles of the objects, which allows the
spatial profile fitting technique to be applied to the space-based images as
well as the ground-based images.  Second, we applied the spatial profile
fitting technique to the individual ground-based images (without combining the
images in each band), which is necessary in order to achieve optimal
sensitivity given that the PSFs of the ground-based images vary significantly
from image to image.  In comparison with conventional methods, the new
technique provides higher signal-to-noise ratio measurements, and in contrast
with conventional methods, the new technique accounts for uncertainty
correlations between nearby, overlapping neighbors.

  First, we determined the PSFs of the space- and ground-based images.  For the
space-based images, we approximated the PSFs by the median average of the three
faintest of the four brightest stars in each image.  For the ground-based
images, we approximated the PSFs by double Gaussian profiles
\begin{equation}
\Phi\left( r \right) =
\sum _{i=1}^2{A_i \exp \left[ -\frac{1}{2}\left(
\frac{r}{\sigma _i}\right)^2\right]}
\end{equation}
(where $r$ is the distance from the profile center), where we estimated
parameters by fitting to the brightest stars in each image (excluding saturated
stars).

  Next, we modeled the intrinsic spatial profiles of the objects.  We produced
the models by reconstructing the F160W and F222M images using the non-negative
least-squares (NNLS) image reconstruction method \citep[see][and references therein]
{Puetter99} and masking the reconstructed image with the SExtractor
segmentation map to identify individual intrinsic spatial profiles of
individual objects.  Briefly, the NNLS image reconstruction method is an
indirect image reconstruction method that constrains the reconstructed image to
be non-negative, which forces the reconstructed image---i.e.\ the model image
convolved with the PSF---to be smooth on the scale of the PSF.  The NNLS image
reconstruction method is matched to our purpose of modeling the intrinsic
spatial profiles of the objects because it produces a smooth model of an image
with the effects of the PSF removed.

  Next, we formed image templates of the objects in the images by convolving
the intrinsic spatial profiles of the objects with the appropriate PSFs of the
images.

  Finally, we fitted the image templates to the images to determine optimal
flux estimates.  For the $k$th image of a given band, we calculated a $\chi^2$
statistic of the form
\begin{equation}
\chi ^2_k = \sum_{i,j} \left[ \frac{I^{(k)}(i,j) - B^{(k)}(i,j)
- \sum_{n = 1}^N F_n P_n^{(k)}(i,j)} {\sigma^{(k)}_{\rm eff}(i,j)} \right]^2,
\end{equation}
where $I^{\left( k\right)}$ is the image, $B^{\left( k\right)}$ is the
background, $P_n^{\left( k\right)}$ is the image template, $\sigma^{(k)}_{\rm
eff}$ is the effective uncertainty (described previously by FLY99) and $F_n$ is
the optimal flux estimate of the $n$th object and where the sum extends over
all pixels in the image.  We determined local backgrounds by median averaging
within $64 \times 64$ pixel$^2$ boxes centered on the objects, excluding pixels
occupied by any objects, and we determined effective uncertainties by summing
the elements of $3 \times 3$ data covariance matrices.  We formed the total
$\chi^2$ of a given band by summing the $\chi^2$ statistic over all individual
images, i.e.\
\begin{equation}
\chi ^2 = \sum _{k}{\chi ^2 _k}.
\end{equation}
For the space-based images, only one image (i.e.\ the final processed image)
enters into the sum, whereas for the ground-based images, $\approx 20$ images
(i.e.\ the individual exposures) enter into the sum.  Given $N$ objects
detected in the image, we set $\partial{\chi ^2} / \partial{F_i} = 0$ to yield
a system of $N$ coupled linear equations with $N$ unknowns (i.e.\ the $F_i$,
with $i = 1,N$).  We solved the equations by Cholesky decomposition of the
Hessian matrix to determine the optimal flux estimates $F_i$ and the optimal
flux uncertainty estimates $\sigma_{F_i}$.  Note that the technique is
applicable to a set of unadded images simply because the Hessian matrix of a
given band is additive with respect to the individual images.

\begin{figure}
\label{fig:snr}
\centerline{\vbox{\plotone{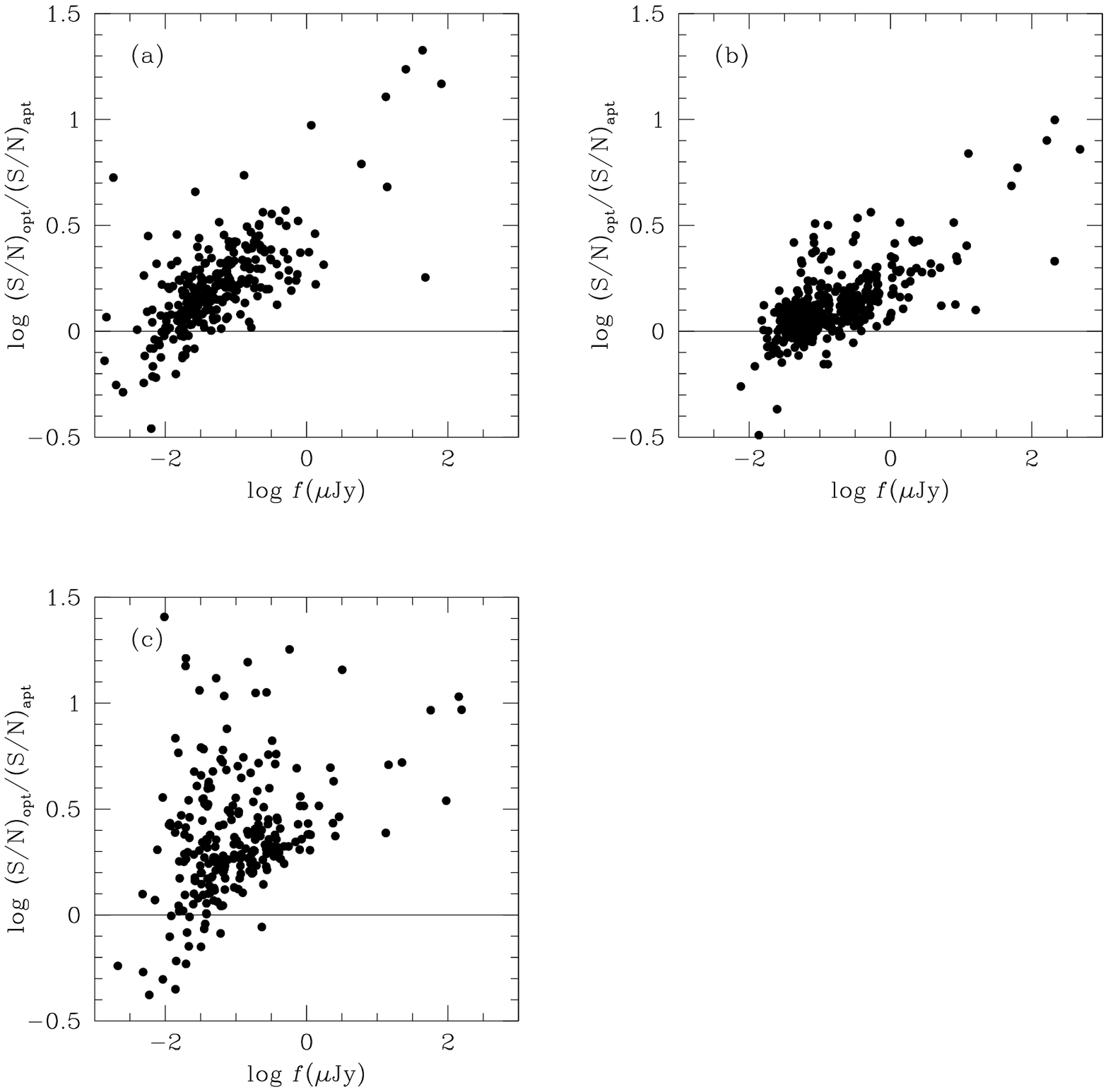}}}
\caption{
Comparison of signal-to-noise ratios obtained by quasi-optimal photometry
method and by direct integration within isophotal apertures versus logarithm of
energy flux obtained by quasi-optimal photometry method for (a) STIS image,
(b) NICMOS F160W image, and (c) VLT {\it I\/}-band image.  The signal-to-noise
ratios obtained by the quasi-optimal photometry method are in general larger
than the signal-to-noise ratios obtained by direct integration within isophotal
apertures.}
\end{figure}

  The signal-to-noise ratios obtained by the current method are in general
substantially larger than the signal-to-noise ratios obtained by conventional
methods, say by direct integration within isophotal apertures.  This is
demonstrated in \reffig{fig:snr}, which shows the signal-to-noise ratio obtained
by the current method compared with the signal-to-noise ratio obtained by the
aperture method versus object flux, for objects measured in the F160W, STIS,
and VLT {\it I\/}-band images.  For the majority of objects at all flux levels,
the signal-to-noise ratio obtained by the current method is larger than the
signal-to-noise ratio obtained by the aperture method, by a factor that is
typically $\approx 2$ but that ranges up to $\approx 10$.  The improvement is
particularly substantial for the ground-based image where the PSFs vary
significantly from exposure to exposure.  For a small minority of objects,
the signal-to-noise ratio obtained by the current method is formally smaller
than the signal-to-noise ratio obtained by the aperture method.  This is
explained by noting that the current method accounts for uncertainty correlations
between nearby, overlapping neighbors whereas the aperture method does not.
In particular, by examining individual objects with low signal-to-noise ratio
detections, we find that: (1) objects with overlapping neighbors of comparable
flux levels have flux errors underestimated by aperture photometry due to
significant contributions from the off-diagonal parts of the covariance matricies,
and (2) objects with overlapping neighbors of much higher flux levels have flux
measurements overestimated by aperture photometry because substantial amounts
of flux from the brighter objects are incorrectly assigned to the fainter objects.
These two errors of aperture photometry have the effect of falsely increasing
the signal-to-noise ratio of the aperture measurements at low signal-to-noise
ratios.  We therefore conclude that the current method is superior
to conventional methods in two respects:  (1)  it provides substantially higher
signal-to-noise ratios, and (2) it provides more realistic error estimates.

\subsection{Photometric Redshifts}

  We measured photometric redshifts following procedures similar to those
described previously by LYF96 and FLY99, but with a sequence of six
spectrophotometric templates, including the four templates of our previous
analyses (of E/S0, Sbc, Scd, and Irr galaxies) and two new templates (of
star-forming galaxies).

\begin{figure}
\label{fig:sed}
\centerline{\vbox{\plotone{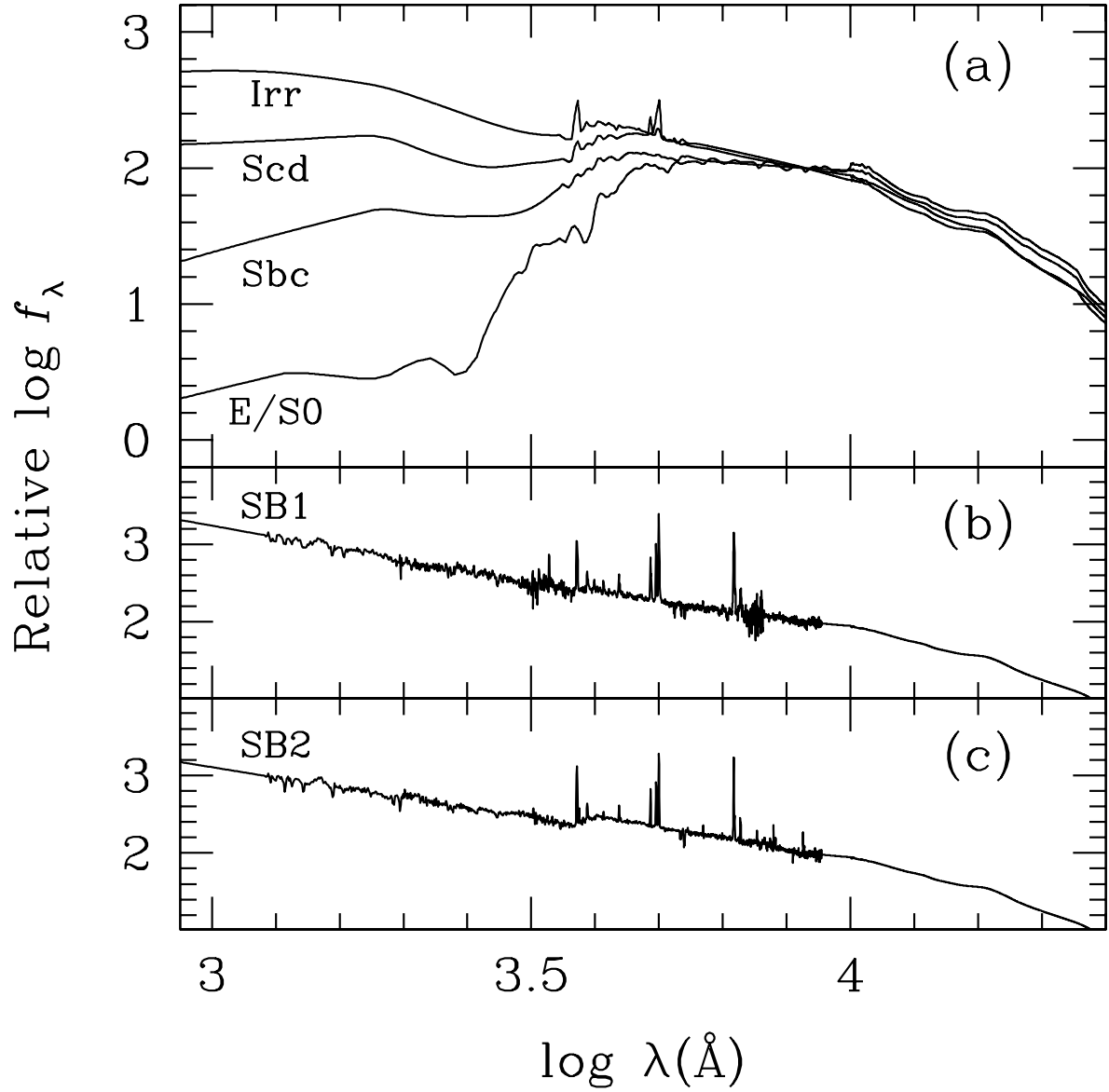}}}
\caption{
Spectral energy distributions of (a) E/S0, Sbc, Scd, and Irr galaxies, (b)
SB1 galaxy, and (c) SB2 galaxy.}
\end{figure}

  First, we formed the two new spectrophotometric templates of star-forming
galaxies by adopting ultraviolet- and optical-wavelength spectrophotometry of
starburst galaxies of intrinsic color excess $E_{B-V} < 0.10$ (designated SB1)
and $0.11 \leq E_{B-V} \leq 0.21$ (designated SB2) of \citet{Kinney96} and
extrapolating toward ultraviolet and infrared wavelengths and incorporating the
effects of intrinsic and intervening absorption according to the prescription
of FLY99.  (Specifically, we assumed that galaxies are optically thick at the
Lyman Limit and incorporated the average Lyman $\alpha$ and Lyman $\beta$
decrement parameters of \citealt{Madau95} and \citealt{Webb96}.)  \reffig{fig:sed}
shows the spectrophotometric templates, which span rest-frame wavelengths
$912 - 25,000$ \AA.

\begin{figure}
\label{fig:stf}
\centerline{\vbox{\plotone{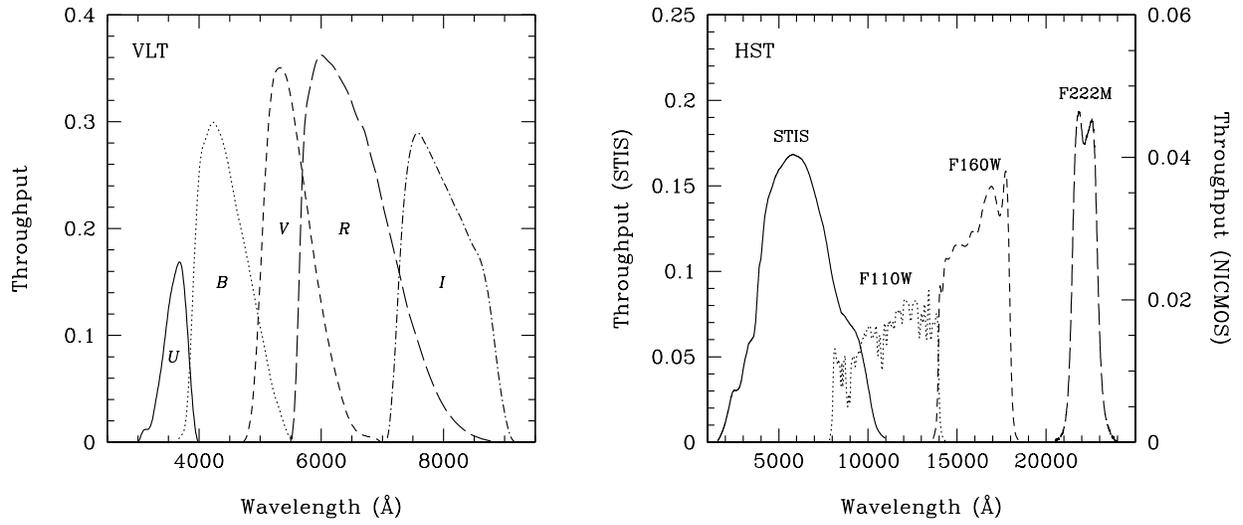}}}
\caption{
System throughput functions of (a) the VLT with the Test Camera and the {\it U},
{\it B}, {\it V}, {\it R}, and {\it I\/} filters and (b) HST with STIS and with
NICMOS and the F110W, F160W, and F222M filters.}
\end{figure}

  Next, we integrated the spectrophotometric templates with the system
throughput functions of each instrument with each filter.  For the HST
instruments, we adopted the system throughput functions provided by the STScI
NICMOS and STIS teams.  For the VLT instruments, we modeled the system
throughput functions using filter and detector response functions provided by
the VLT Science Verification team, instrument response functions calculated from
the measured reflectivity of the Al reflecting surfaces, and a standard atmospheric
response function.  \reffig{fig:stf} shows the system throughput functions.

  Finally, we determined photometric redshifts by maximizing a likelihood
estimator of the form
\begin{equation}
{\cal L}\left( z, T \right) = \prod_{i=1}^{9}{\exp \left\{ -\frac{1}{2}
\left[ \frac{f_i - A F_i \left( z, T \right)} {\sigma_i} \right]^2 \right\}},
\end{equation}
where $f_i$ is the measured flux in band $i$, $\sigma_i$ is the measured flux
uncertainty in band $i$, $F_i(z,T)$ is the modeled flux in band $i$ at assumed
redshift $z$ and spectral type $T$, and $A$ is an arbitrary flux normalization
and where the product extends over all nine bands.  For each object, ${\cal
L}(z,T)$ was maximized with respect to $A$ and $T$ to determine the ``redshift
likelihood function'' ${\cal L}(z)$, which was maximized with respect to $z$ to
determine the maximum-likelihood photometric redshift.

\subsection{Star and Galaxy Separation}

  We identified probable stars on the basis of visual inspection of the
space-based images and the spectral energy distributions.  Although in
principle stars might be identified based solely on their spectrophotometric
characteristics, in practice stars occur so infrequently in the HDF or HDF--S
images that we decided not to incorporate stellar spectrophotometric templates
into the photometric redshift analysis.  A total of five probable stars were
identified, the brightest of which is of magnitude $AB(16,000) \approx 17.2$
and the faintest of which is of magnitude $AB(16,000) \approx 21.1$.
\reftbl{tbl:stars} lists properties of the probable stars.

\section{RESULTS}

\subsection{Catalog of Photometry and Photometric Redshifts}

  The result of the analysis described in the previous section is a catalog of
photometry and photometric redshifts of 335 objects in the HDF--S NICMOS field.
The catalog is available on a World Wide Web site at {\tt
http://\-www.\-ess.\-sunysb.\-edu/\-astro/\-hdfs/\-index.\-html}.  For each
object, the catalog lists (1) the object identification, (2--3) the J2000 right
ascension and declination, (4) the F160W magnitude $AB(16,000)$, (5--22) the
relative energy flux density per unit frequency interval and uncertainty in the
{\it U}, {\it B}, {\it V}, {\it R}, {\it I}, STIS, F110W, F160W, and F222M
bands (with respect to the F160W band), (23) the best-fit photometric redshift,
and (24) the best-fit spectral type. Here $AB$ magnitude is related to energy
flux density per unit frequency interval $f_\nu$ as
\begin{equation}
AB(\lambda) = -2.5 \log{\frac{f_\nu(\lambda)}{1\;\mu {\rm Jy}}} + 23.90.
\end{equation}
The World Wide Web site also includes individual object pages that display the
redshift likelihood functions, measured and modeled spectral energy
distributions, and images of the objects.

\subsection{Survey Area versus Depth Relation}

\begin{figure}
\label{fig:area}
\centerline{\vbox{\plotone{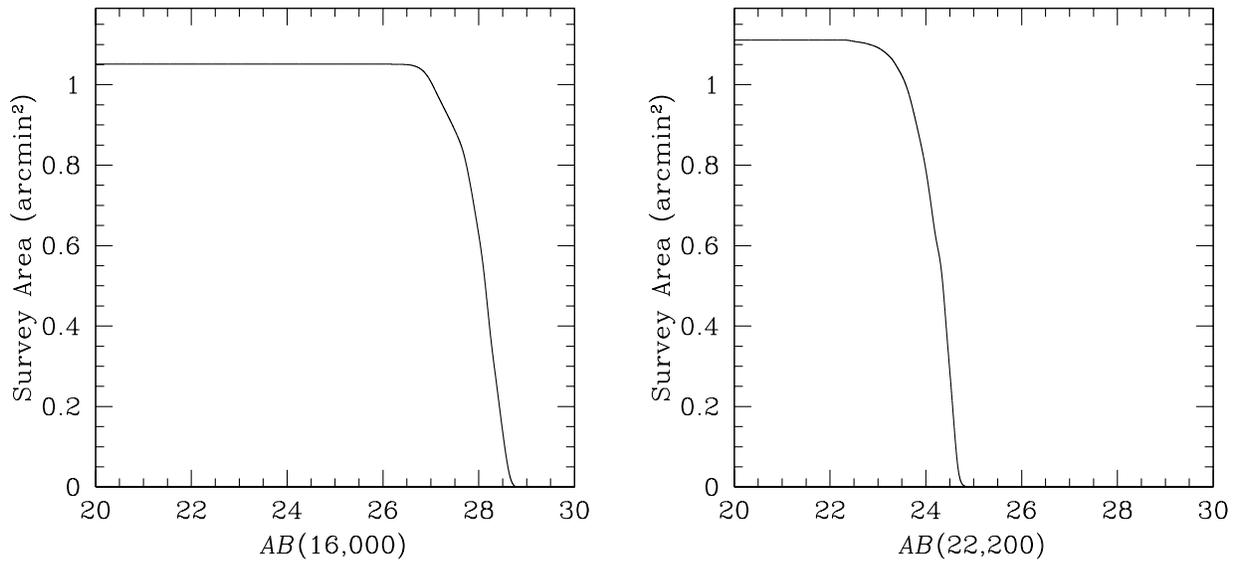}}}
\caption{
Survey area versus depth relation on the basis of (a) NICMOS F160W image and
(b) NICMOS F222M image.}
\end{figure}

  The primary utility of the catalog of photometric redshifts is as a survey of
faint galaxies detected in the NICMOS F160W image.  Because the sensitivity of
the F160W image varies significantly with position, the selection function of
the survey is characterized by the survey area versus depth relation.  We
determined the survey area versus depth relation by assuming that the
sensitivity image traces the shape (but not necessarily the normalization) of
the sensitivity versus position relation of the F160W image.  (The sensitivity
image may not trace the normalization of the sensitivity versus position
relation because of the non-zero off-diagonal elements of the covariance
matrix.)  First, we formed the sensitivity image by taking the square root of
the variance image.  Next, we scaled the sensitivity image so that it traced
the faint-end envelope of the measured brightnesses of objects detected in the
F160W image.  Finally, we integrated the enclosed area as a function of
limiting depth to determine the survey area versus depth relation.
\reffig{fig:area} shows the survey area versus depth relation, which indicates
that the survey reaches a limiting depth of $AB(16,000) \approx 28.7$ and covers
1.01 arcmin$^2$ to $AB(16,000) = 27$ and 1.05 arcmin$^2$ to $AB(16,000) = 26.5$.
Likewise, the survey reaches a limiting depth of $AB(22,200) \approx 24.8$ and
covers 0.79 arcmin$^2$ to $AB(22,200) = 24$ and 1.09 arcmin$^2$ to $AB(22,200)
= 23$.  The survey area versus depth relation is crucial to any statistical
analysis of the catalog.

\section{EVALUATION OF THE PHOTOMETRIC REDSHIFT TECHNIQUE}

\subsection{Accuracy and Reliability of the Photometric Redshift Measurements}

  Spectroscopic redshift measurements of $\approx 120$ faint galaxies in the HDF
have been obtained using the Keck telescope (see, e.g., the list compiled by
FLY99), and spectroscopic measurements of three galaxies in the HDF--S WFPC2
field have been obtained using the Anglo-Australian Telescope
\citep[in preparation]{Glazebrook00}.  Although more such measurements will
undoubtedly be obtained (especially of faint galaxies in the HDF--S), the current
measurements provide a means of assessing the accuracy and reliability of the
photometric redshift measurements and of comparing results of the four- versus
six-template photometric redshift measurements.

\begin{figure}
\label{fig:comp}
\centerline{\vbox{\plotone{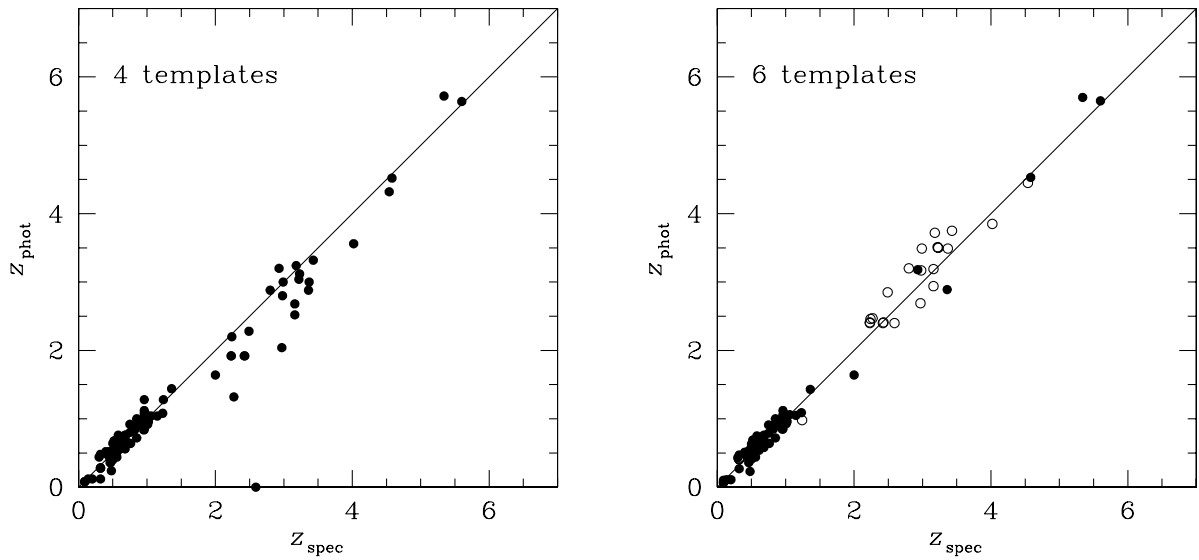}}}
\caption{
Comparison of 104 photometric and reliable spectroscopic redshift measurements
for (a) four-template photometric redshift measurements and (b) six-template
photometric redshift measurements.  Filled circles represent E/S0, Sbc, Scd,
and Irr best-fit spectral types, and open circles represent SB1 and SB2
best-fit spectral types.}
\end{figure}

  We compiled spectroscopic redshift measurements from published and
unpublished sources, rejecting as unreliable spectroscopic measurements with
uncertain or ambiguous interpretations.  [A non-negligible fraction of the
spectroscopic redshift measurements have been shown to be in error and so must
be excluded from consideration; see, e.g., the discussions of \citet[hereafter LFY98]
{LFY98} and FLY99.]  \reffig{fig:comp} shows the comparison of 104 photometric
and reliable spectroscopic redshift measurements.  Specifically, \reffig{fig:comp}(a)
shows the comparison of the four-template photometric redshift measurements with the
reliable spectroscopic redshift measurements, and \reffig{fig:comp}(b) shows the
comparison of the six-template photometric redshift measurements with the reliable
spectroscopic redshift measurements.  Several results are evident on the basis
of \reffig{fig:comp}:

  1.  Inclusion of the two new templates eliminates the tendency of our
previous analyses to systematically underestimate the redshifts of galaxies of
redshift $2 < z < 3$ (by a redshift offset of roughly 0.3), in agreement with
results found previously by \citet{Benitez98}.  The six-template photometric
redshift measurements are essentially free of systematic bias at all redshifts
$z < 6$ that have as yet been examined.

  2.  The RMS residual between the six-template photometric redshift
measurements and the reliable spectroscopic redshift measurements is $\Delta z
= 0.09$ at redshifts $z < 2$, $\Delta z= 0.29$ at redshifts $2 < z < 4$, and
$\Delta z = 0.18$ at redshifts $z > 4$.  The median absolute residual between
the six-template photometric redshift measurements and the reliable spectroscopic
redshift measurements is $\Delta z = 0.07$ at redshifts $z < 2$, $\Delta z= 0.22$
at redshifts $2 < z < 4$, and $\Delta z = 0.09$ at redshifts $z > 4$.

  3.  The six-template photometric redshift measurements are accurate to within
an RMS relative uncertainty of $\Delta z/(1 + z) \lesssim 7\%$ at all redshifts $z
< 6$ that have as yet been examined.

  We conclude that the photometric redshift technique is in general capable of
determining reliable redshifts to within a relative uncertainty of $\Delta z/(1
+ z) \lesssim 7\%$. 

\subsection{Photometric Redshift Measurements of Stars}

  The photometric redshift measurements of the probable stars listed in \reftbl{tbl:stars}
are $z = 0.07$, 0.30, 5.33, 5.63, and 5.72.  Thus the spectral energy distributions of 
some stars resemble the spectral energy distributions of galaxies of redshift
$z = 5 - 6$.  We believe that all such stars were identified on the basis of visual
inspection of the space-based images, but we cannot exclude the possibility that a
small number of faint stars were misidentified as galaxies of redshift $z = 5 - 6$.

\subsection{Effects of Photometric Error on the Photometric Redshift
Measurements}

  Comparison of the photometric and spectroscopic redshift measurements yields
a measure of the uncertainties of the photometric redshift technique, which in
principle can include contributions from both photometric error and cosmic
variance with respect to the spectral templates.  At bright magnitudes, the
effects of photometric error are expected to be negligible, while at faint
magnitudes the effects of photometric error are expected to dominate.

\begin{figure}
\label{fig:sim}
\centerline{\vbox{\plotone{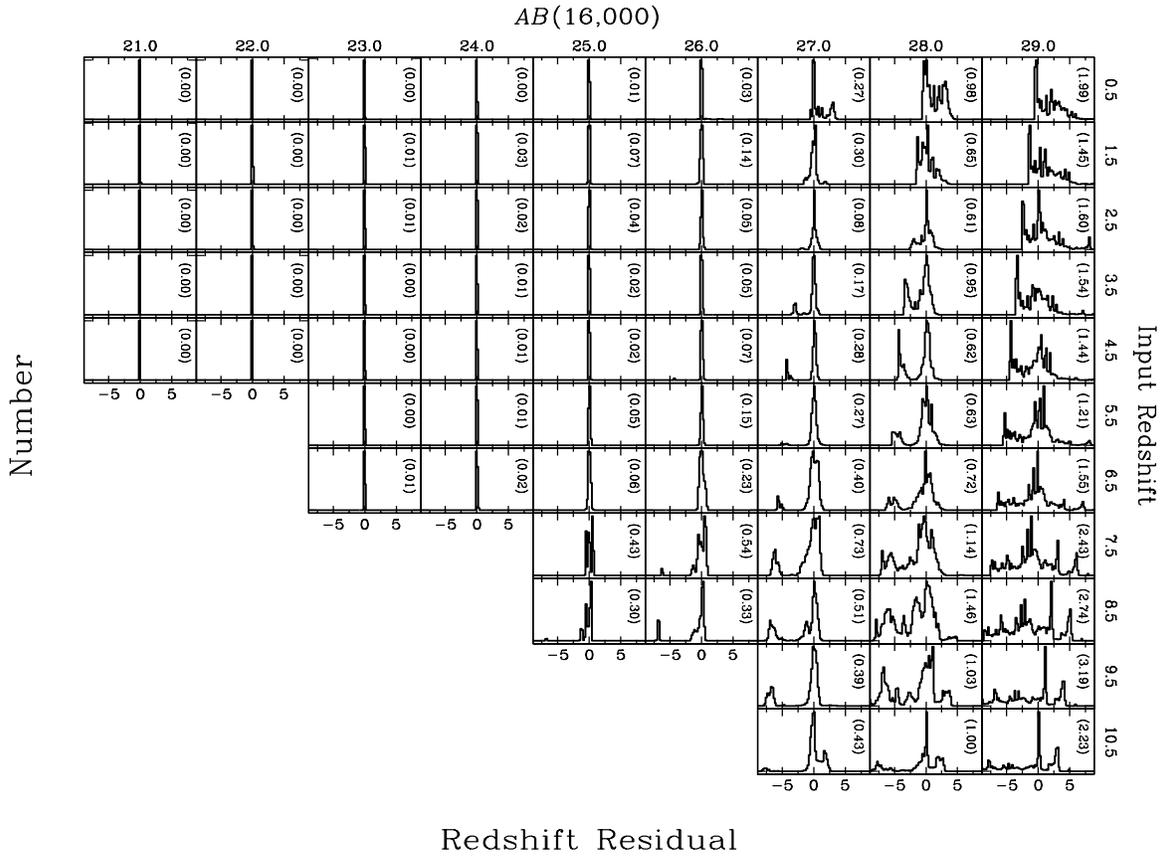}}}
\caption{
Distribution of redshift residuals as functions of galaxy magnitude
$AB(16,000)$ and redshift $z$.  The number in parentheses gives the median
absolute deviation of the output redshift from the input redshift for each
panel.}
\end{figure}

  We assessed the effects of photometric error on the photometric redshift
measurements by performing a series of simulations similar to those described
previously by LFY98.  First, we determined the expected energy fluxes
through the various filters of an Irr galaxy spectrophotometric template, given
an assumed galaxy magnitude $AB(16,000)$ [selected over the range
$21 < AB(16,000) < 29$] and redshift $z$ (selected over the range $0 < z < 11$).
Next, we added random noise to the expected energy fluxes according to the actual
noise characteristics of the images.  Next, we determined photometric redshift
measurements of the simulated objects using the sequence of six spectrophotometric
templates.  Finally, we repeated these steps 1000 times as functions of $AB(16,000)$
and $z$ to determine the distribution of redshift residuals between the input and
output models.  \reffig{fig:sim} shows the distributions of redshift residuals as
functions of $AB(16,000)$ and $z$.  Several results are evident on the basis of
\reffig{fig:sim}:

  1. At $AB(16000) < 25$, photometric errors have a negligible effect on the
photometric redshift measurements.  At these relatively bright magnitudes, the
RMS dispersion of the residuals is $\Delta z \lesssim 0.02$.

  2. At $AB(16000) = 25 - 26$, photometric errors have only a modest effect on
the photometric redshift measurements at redshifts $z \lesssim 7$, where the RMS
dispersion of the residuals is $\Delta z \lesssim 0.25$, but have a somewhat more
significant effect on the photometric redshift measurements at $z \gtrsim 7$,
where a secondary peak in the residual distribution occurs at large negative
residual, i.e.\ at $\Delta z \approx -6$.  The secondary peak is caused by
ambiguity between high-redshift late-type galaxies and low-redshift early-type
galaxies.

  3. At $AB(16000) = 27 - 28$, photometric errors have a modest effect on the
photometric redshift measurements at all redshifts, with a prominent secondary
peak in the residual distribution at all redshifts $z \gtrsim 3$.  The sense of
the secondary peak is such that it is more likely for high-redshift objects to
be assigned low redshifts than for low-redshift objects to be assigned high
redshifts.

  4. At $AB(16000) > 28$, photometric errors have a significant effect on the
photometric redshift measurements at all redshifts.

  We conclude that the effects of photometric error on the photometric redshift
measurements must be taken into consideration at magnitudes fainter than
$AB(16,000) \approx 25$.

\section{DISCUSSION}

  Here we briefly discuss results of the catalog of photometric redshifts,
concentrating on results related to the highest-redshift galaxies identified by
the analysis.  Scientific analysis of the catalog will be presented in
forthcoming papers.

\subsection{Redshift Distribution of Galaxies in the HDF--S NICMOS Field}

\begin{figure}
\label{fig:hist}
\centerline{\vbox{\plotone{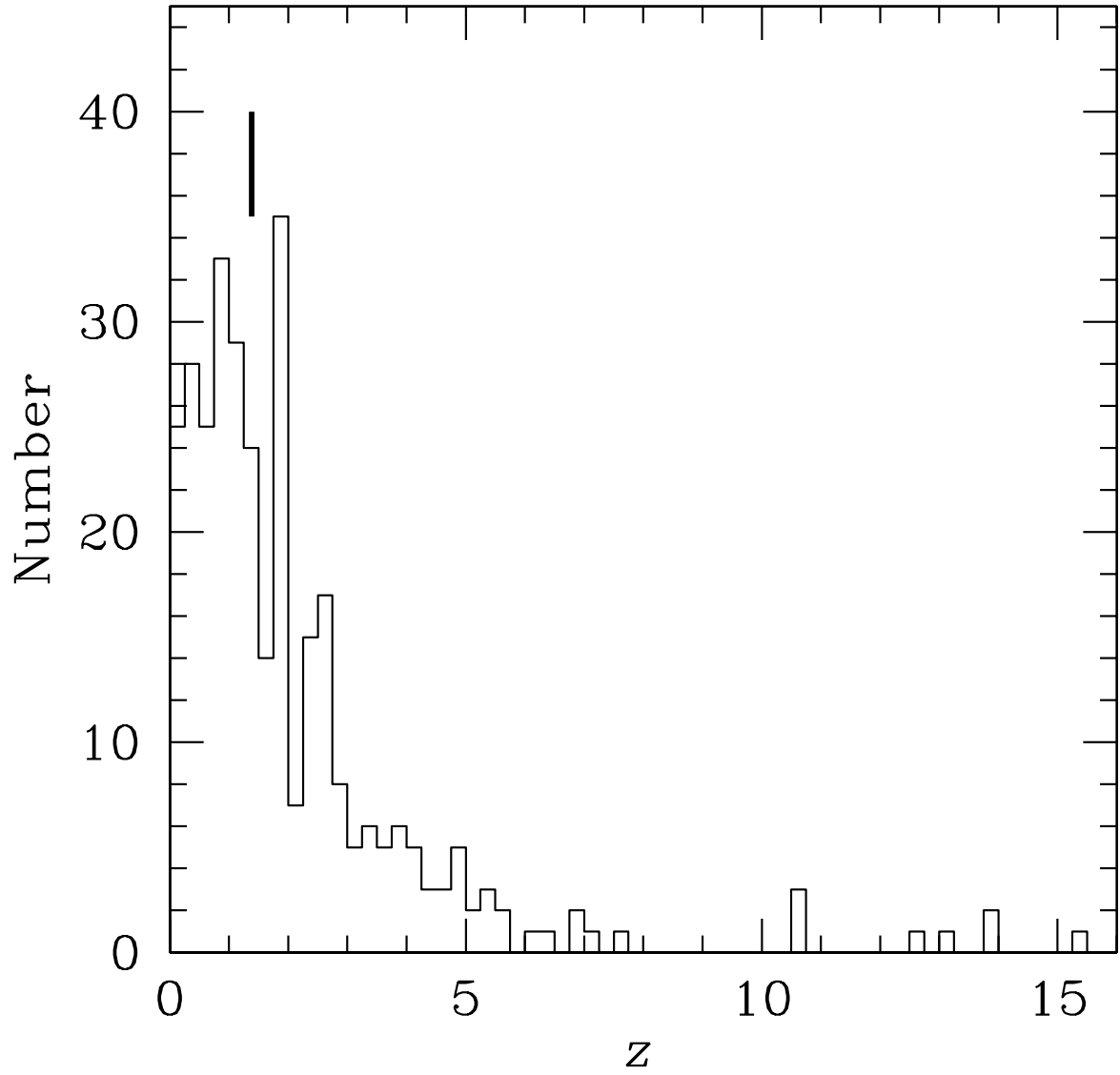}}}
\caption{
Redshift distribution of galaxies identified in the HDF--S NICMOS field.
Line segment indicates median redshift.}
\end{figure}

  The catalog of photometric redshifts identifies 330 galaxies, of photometric
redshift measurement ranging from $z \approx 0$ through $z > 10$.  \reffig{fig:hist}
shows the redshift distribution of the galaxies in the HDF--S NICMOS field.  The
distribution is characterized by a median redshift of $z_{\rm med} = 1.38$ and
by a tail that stretches to redshifts beyond $z = 10$.  The redshift distribution
of \reffig{fig:hist} does not, of course, apply for any magnitude-limited sample,
because the sensitivity of the F160W image varies significantly with position.

\subsection{Galaxies of Redshift $z > 5$}

  One difference between the current analysis of the HDF--S NICMOS field and our
previous analyses of the HDF is that the current analysis is in principle
sensitive to galaxies of redshift larger than was the previous analyses.  In
this section, we discuss the galaxies of photometric redshift measurement $z > 5$.

  The catalog of photometric redshifts identifies 21 galaxies (or 6\% of the
total) of redshift $z > 5$.  \reftbl{tbl:hiz} lists the positions, magnitudes
$AB(16,000)$, photometric redshift measurements $z$, and best-fit spectral
types of these galaxies, and \reffig{fig:hiz} shows the observed and modeled
spectral energy distributions and redshift likelihood functions of these galaxies.
\reftbl{tbl:hizsfd} lists surface densities of galaxies of redshift $z > 5$
derived from the catalog of photometric redshifts accounting for the variation of
the survey area versus depth relation as a function of limiting magnitude $AB(16,000)$.
Uncertainties listed in \reftbl{tbl:hizsfd} are derived by a bootstrap resampling
technique, which explicitly accounts for sampling error, photometric error, and cosmic
variance with respect to the spectrophotometric templates.  First, we resampled
the original catalog, allowing the possibility of duplication.  Next, we added
random noise to the flux measurements (according to the actual noise properties
of the images) and redetermined the photometric redshift measurements.  Next,
we added random noise to the photometric redshift measurements (according to the actual
noise properties of the photometric redshift technique, as described in \S\ 5.1).
Next, we measureed galaxy surface densities from the resampled and perturbed
photometric redshift catalog.  Finally, we repeated these steps a thousand
times and determine the 1 $\sigma$ deviations of the surface density measurements.
We conclude that galaxies of redshift $z > 5$ are a non-negligible fraction of the
galaxy population at magnitudes $AB(16,000) \gtrsim 27$.

\subsection{Galaxies of Redshift $z > 10$}

  The catalog of photometric redshifts identifies 8 galaxies (or 2\% of the total)
of redshift $z > 10$ including 3 galaxies detected on the basis of the F222M image.
\reftbl{tbl:hizsfd2} lists surface densities of galaxies of redshift $z > 10$
derived from the catalog of photometric redshifts accounting for the variation of
the survey area versus depth relation as a function of limiting magnitudes
$AB(16,000)$ and $AB(22,200)$.  We are struck that the surface density of galaxies
of redshift $z > 10$ at $AB(22,200)=24$ is comparable to the surface density of
galaxies of redshift $z > 10$ at $AB(16,000)=28$.

\subsection{Early-Type Galaxies of Redshift $z > 1$}

  Another difference between the current analysis of the HDF--S NICMOS field and
our previous analyses of the HDF is that the current analysis is in principle
sensitive to early-type galaxies of redshift larger than was the previous
analyses.  In this section, we discuss the early-type galaxies of photometric
redshift measurement $z > 1$.

  The catalog of photometric redshifts identifies 11 galaxies (or 3\% of the
total) of best-fit spectral type E/S0, of which 5 galaxies (or 1\% of the
total) are of redshift $z > 1$.  \reftbl{tbl:hie} lists the positions, magnitudes
$AB(16,000)$, photometric redshift measurements $z$, and \reffig{fig:hie} shows
the observed and modeled spectral energy distributions and redshift likelihood
functions of these galaxies.  (It should be noted, however, that the likelihood
that SB--NI--0471--0941 is an early--type galaxy of redshift $z = 5.10$ is
statistically indistinguishable from the likelihood that this object is a
later--type galaxy of redshift $z \gtrsim 10$.  A similar result also applies
for SB--NI--0844--0698.)

\reftbl{tbl:hiesfd} lists surface densities of early-type galaxies of redshift
$z > 1$ derived from the catalog of photometric redshifts and the survey area
versus depth relation as a function of limiting magnitude $AB(16,000)$.
(Uncertainties listed in \reftbl{tbl:hiesfd} are derived as in \reftbl{tbl:hizsfd}.)
We conclude that early-type galaxies of redshift $z > 1$ are a non-negligible
fraction of the galaxy population at magnitudes $AB(16,000) \gtrsim 25$.

\subsection{Comparison with Other Photometric Redshift Measurements}

\reftbl{tbl:comp} compares photometric redshift measurements of six galaxies in
common between the current analysis and the previous analysis of \citet{Benitez99}.
Four of the six pairs of measurements are concordant and two of the six pairs
of measurements are discordant to within the cosmic dispersion $\Delta z \approx
0.1$ of the photometric redshift measurement technique as described in \S\ 5.1.

\citet{Benitez99} used similar photometric redshift techniques and template
spectra, so the discrepancies most likely arise from differences in photometry.
Comparison of photometry in all bands indicates that, while our flux
measurements are consistent with the flux measurements of \citet{Benitez99} in
the optical and the F160W bands, our flux measurements are systematically lower
in the F110W band and systematically higher in the F222M band.  These
differences arise because we use a spatial profile fitting technique, which
takes into account differences in the PSF from band to band, whereas
\citet{Benitez99} use a fixed aperture technique (where the apertures are
determined from a combined F110W and F160W image), which implicitly assumes
identical PSFs in all bands.  There are indeed differences in the PSFs for the
three NICMOS images.  The PSF of the F222M image is slightly broader than the
PSF of the F160W image, and the PSF of the F110W image is slightly sharper than
the PSF of the F160W image.  We therefore suspect that \citet{Benitez99} may
have overestimated fluxes in the F110W image (because the apertures were too
large) and underestimated fluxes in the F222M image (because the apertures were
too small).

\section{SUMMARY AND CONCLUSIONS}

  Here we present a catalog of photometry and photometric redshifts of 335
faint objects in the HDF--S NICMOS field.  The analysis is based on (1) infrared
images obtained with HST using NICMOS with the F110W, F160W, and F222M filters,
(2) an optical image obtained with HST using STIS with no filter, and (3) optical
images obtained with the ESO VLT with {\it U}, {\it B}, {\it V}, {\it R}, and {\it I\/}
filters.  The primary utility of the catalog of photometric redshifts is as a
survey of faint galaxies detected in the NICMOS F160W and F222M images.  The
sensitivity of the survey varies significantly with position, reaching a limiting
depth of $AB(16,000) \approx 28.7$ and covering 1.01 arcmin$^2$ to $AB(16,000) = 27$
and 1.05 arcmin$^2$ to $AB(16,000) = 26.5$. Likewise, the survey reaches a limiting
depth of $AB(22,200) \approx 24.8$ and covering 0.79 arcmin$^2$ to $AB(22,200) = 24$
and 1.09 arcmin$^2$ to $AB(22,200) = 23$.  The catalog of photometric redshifts
identifies 21 galaxies (or 6\% of the total) of redshift $z > 5$, 8 galaxies
(or 2\% of the total) of redshift $z > 10$, and 11 galaxies (or 3\% of the total)
of best-fit spectral type E/S0, of which 5 galaxies (or 1\% of the total) are of
redshift $z > 1$.

\acknowledgements

  The authors thank Bob Williams and the entire STScI HDF--S team for providing
access to the HDF--S observations, the entire ESO VLT team for making observations
of the HDF--S publicly available, and an anonymous referee for helpful comments.
HWC, KML, SMP, and NY were supported by NASA grant NAGW--4422 and NSF grant
AST--9624216.  AFS was supported by a grant from the Australian Research Council.
RCP was supported by NASA grant NAG--53944.  AY was supported by NASA grant
AR--07551.01--96A.


\clearpage

\clearpage

\clearpage

\clearpage

\clearpage

\clearpage

\clearpage

\clearpage
\begin{figure}
\label{fig:hiz}
\centerline{\vbox{\plotone{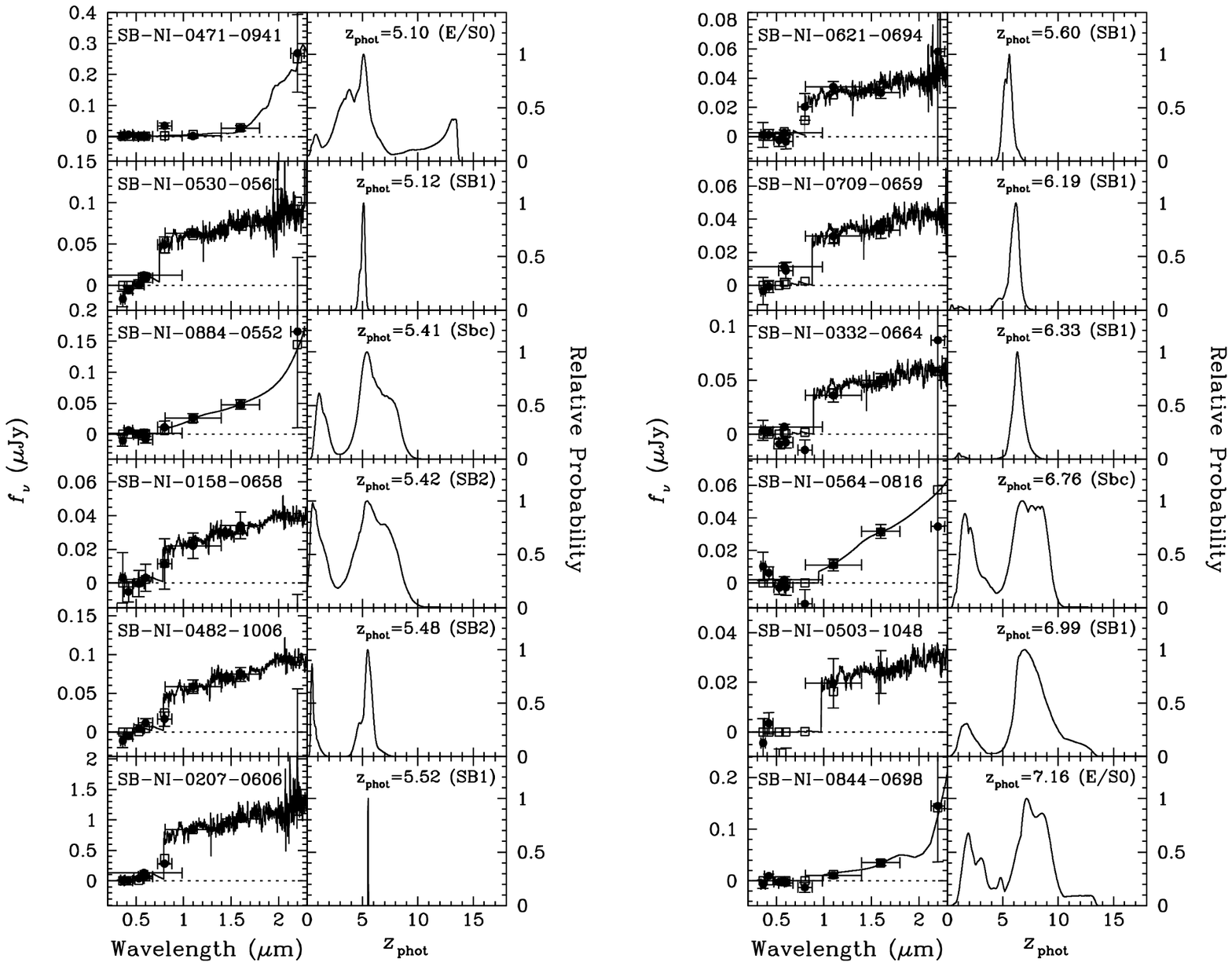}}}
\caption{
Observed and modeled spectral energy distributions (left panels) and redshift
likelihood functions (right panels) of galaxies of photometric redshift
measurement $z > 5$.  Filled circles are measured fluxes and open squares
are best-fit model fluxes.  Vertical error bars indicate $1 \sigma$ uncertainties
and horizontal error bars indicate filter FWHM.  Photometric redshift measurement
and best-fit spectral type of each galaxy are indicated.}
\end{figure}

\clearpage
\addtocounter{figure}{-1}
\begin{figure}
\centerline{\vbox{\plotone{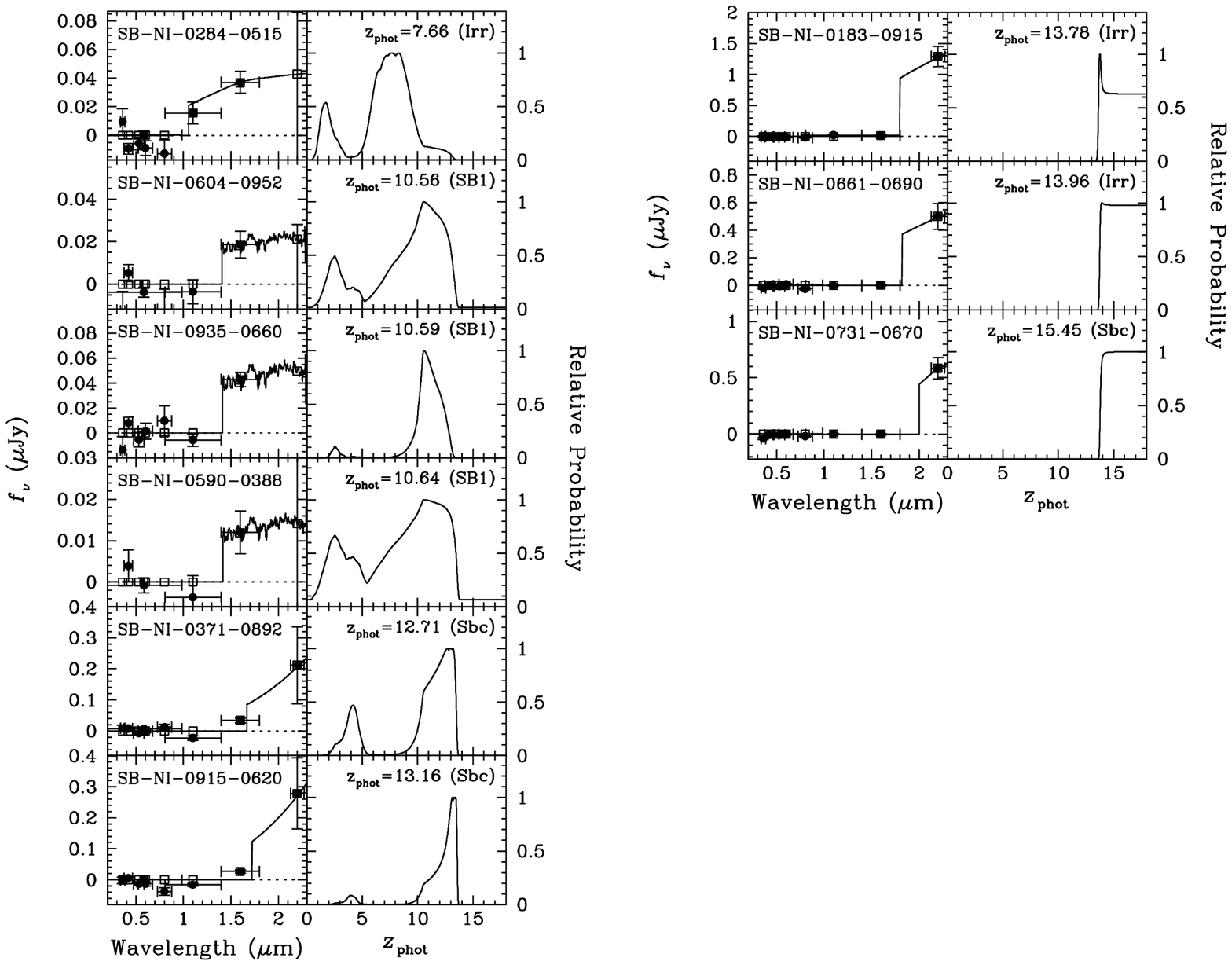}}}
\caption{--- {\it Continued.}}
\end{figure}

\clearpage
\begin{figure}
\label{fig:hie}
\centerline{\vbox{\plotone{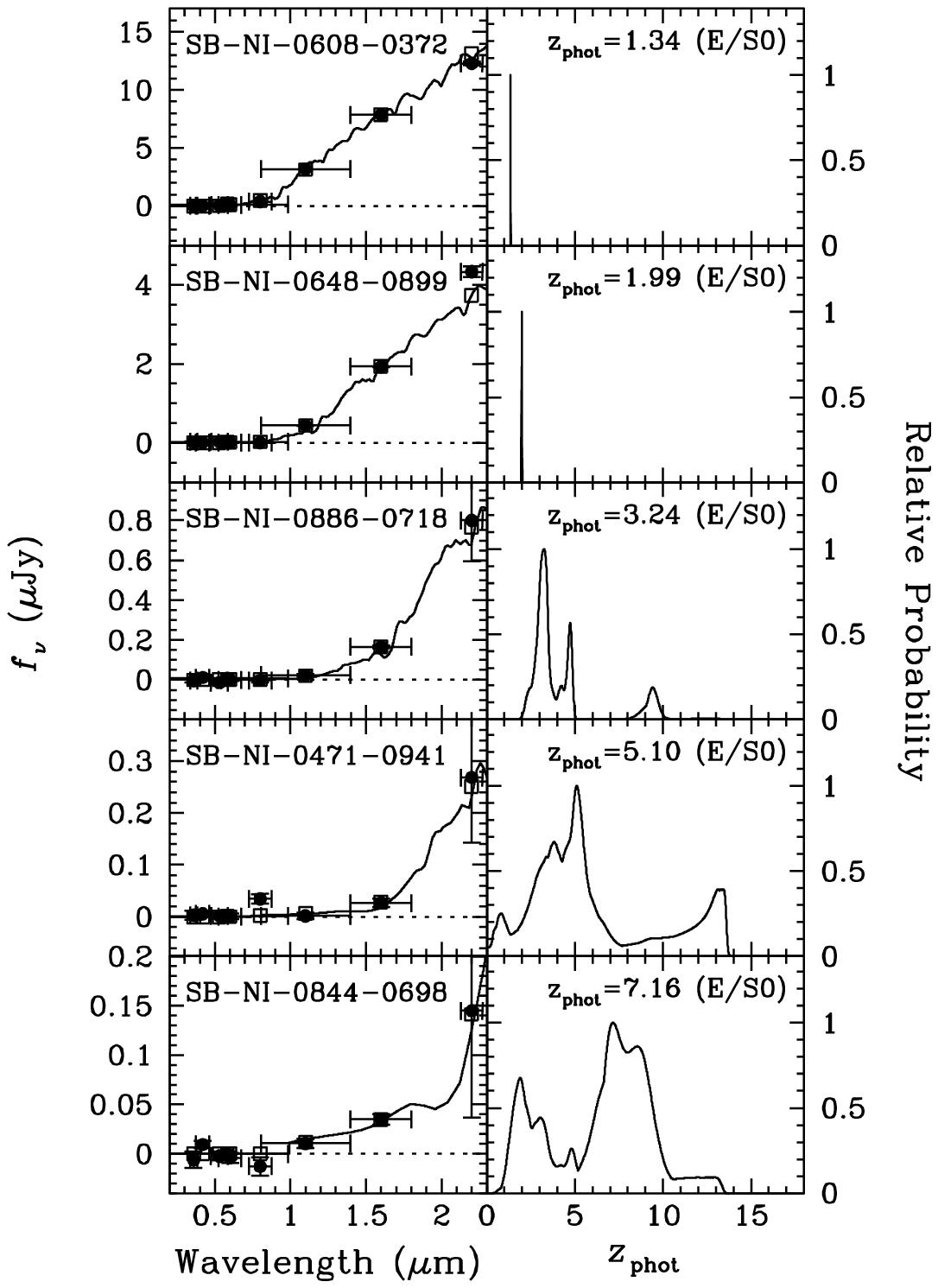}}}
\caption{
Observed and modeled spectral energy distributions (left panels) and redshift
likelihood functions (right panels) of early-type galaxies of photometric
redshift measurement $z > 1$.  Filled circles are measured fluxes and open
squares are best-fit model fluxes.  Vertical error bars indicate $1 \sigma$
uncertainties and horizontal error bars indicate filter FWHM.  Photometric
redshift measurement and best-fit spectral type of each galaxy are indicated.}
\end{figure}

\clearpage

\begin{deluxetable}{p{2.5in}crrr}
\tablewidth{0pt}
\tablecaption{
\label{tbl:obs}
Details of the Observations
}
\tablehead{
\colhead{} & 
\colhead{} & 
\colhead{} & 
\colhead{} & 
\colhead{Exposure}
\\
\colhead{Instrument} &
\colhead{Filter} & 
\colhead{$\lambda$ (\AA)} &
\colhead{$\Delta\lambda$ (\AA)} & 
\colhead{Times (s)}
}
\startdata
ESO VLT (UT1) \dotfill       & VLT-TC U      &   3600 &  500 &  32,400 \\
                             & VLT-TC B      &   4200 &  900 &  10,800 \\
                             & VLT-TC V      &   5300 & 1100 &  10,800 \\
                             & SUSI2 \#825 R &   6000 & 1500 &  10,800 \\
                             & VLT-TC I      &   8000 & 1500 &  14,400 \\
HST STIS 50CCD \dotfill      & \nodata       &   5835 & 8000 &  25,900 \\
HST NICMOS Camera 3 \dotfill & F110W         & 10,985 & 5915 & 108,539 \\
                             & F160W         & 15,940 & 4030 & 128,441 \\
                             & F222M         & 22,160 & 1432 & 103,163 \\
\enddata
\end{deluxetable}

\clearpage

\begin{deluxetable}{p{2.0in}ccc}
\tablewidth{0pt}
\tablecaption{
\label{tbl:stars}
Probable Stars
}
\tablehead{
\colhead{Object} &
\colhead{$\alpha$} &
\colhead{$\delta$} &
\colhead{$AB(16,000)$}
}
\startdata
SB--NI--0157--0534 \dotfill & 22:32:55.69 & $-$60:38:57.59 & 21.14 \\
SB--NI--0277--0904 \dotfill & 22:32:54.47 & $-$60:38:29.88 & 17.19 \\
SB--NI--0278--0599 \dotfill & 22:32:54.46 & $-$60:38:52.75 & 18.37 \\
SB--NI--0446--0431 \dotfill & 22:32:52.75 & $-$60:39:05.35 & 18.09 \\
SB--NI--0774--0562 \dotfill & 22:32:49.40 & $-$60:38:55.53 & 19.40 \\
\enddata
\end{deluxetable}

\clearpage

\begin{deluxetable}{p{2.0in}ccccc}
\tablewidth{0pt}
\tablecaption{
\label{tbl:hiz}
Galaxies of Redshift $z > 5$
}
\tablehead{
\colhead{Object} &
\colhead{$\alpha$} &
\colhead{$\delta$} &
\colhead{$AB(16,000)$} &
\colhead{$z_{\rm phot}$} &
\colhead{Type}
}
\startdata
SB--NI--0471--0941 \dotfill & 22:32:52.49 & $-$60:38:27.08 & 27.83 & \phm{1}5.10 & E/S0 \\
SB--NI--0530--0561 \dotfill & 22:32:51.89 & $-$60:38:55.57 & 26.76 & \phm{1}5.12 & SB1  \\
SB--NI--0884--0552 \dotfill & 22:32:48.28 & $-$60:38:56.24 & 27.20 & \phm{1}5.41 & Sbc  \\
SB--NI--0158--0658 \dotfill & 22:32:55.68 & $-$60:38:48.28 & 27.56 & \phm{1}5.42 & SB2  \\
SB--NI--0482--1006 \dotfill & 22:32:52.38 & $-$60:38:22.18 & 26.72 & \phm{1}5.48 & SB2  \\
SB--NI--0207--0606 \dotfill & 22:32:55.18 & $-$60:38:52.17 & 23.88 & \phm{1}5.52 & SB1  \\
SB--NI--0621--0694 \dotfill & 22:32:50.96 & $-$60:38:45.63 & 27.71 & \phm{1}5.60 & SB1  \\
SB--NI--0709--0659 \dotfill & 22:32:50.07 & $-$60:38:48.25 & 27.59 & \phm{1}6.19 & SB1  \\
SB--NI--0332--0664 \dotfill & 22:32:53.91 & $-$60:38:47.82 & 27.15 & \phm{1}6.33 & SB1  \\
SB--NI--0564--0816 \dotfill & 22:32:51.54 & $-$60:38:36.42 & 27.65 & \phm{1}6.76 & Sbc  \\
SB--NI--0503--1048 \dotfill & 22:32:52.16 & $-$60:38:19.05 & 27.94 & \phm{1}6.99 & SB1  \\
SB--NI--0844--0698 \dotfill & 22:32:48.69 & $-$60:38:45.28 & 27.54 & \phm{1}7.16 & E/S0 \\
SB--NI--0284--0515 \dotfill & 22:32:54.40 & $-$60:38:59.01 & 27.48 & \phm{1}7.66 & Irr  \\
SB--NI--0604--0952 \dotfill & 22:32:51.13 & $-$60:38:26.23 & 28.23 & 10.56 & SB1  \\
SB--NI--0935--0660 \dotfill & 22:32:47.75 & $-$60:38:48.14 & 27.32 & 10.59 & SB1  \\
SB--NI--0590--0388 \dotfill & 22:32:51.28 & $-$60:39:08.53 & 28.70 & 10.64 & SB1  \\
SB--NI--0371--0892 \dotfill & 22:32:53.51 & $-$60:38:30.77 & 27.58 & 12.71 & Sbc  \\
SB--NI--0915--0620 \dotfill & 22:32:47.97 & $-$60:38:51.13 & 27.84 & 13.16 & Sbc  \\
SB--NI--0183--0915 \dotfill & 22:32:55.43 & $-$60:38:29.03 & \phm{$^{\dag}$}23.62$^{\dag}$ & 13.78 & Irr  \\
SB--NI--0661--0690 \dotfill & 22:32:50.55 & $-$60:38:45.90 & \phm{$^{\dag}$}24.65$^{\dag}$ & 13.97 & Irr  \\
SB--NI--0731--0670 \dotfill & 22:32:49.84 & $-$60:38:47.40 & \phm{$^{\dag}$}24.48$^{\dag}$ & 15.45 & Sbc  \\
\enddata
\tablenotetext{\dag}{$AB(22,200)$}
\end{deluxetable}

\clearpage

\begin{deluxetable}{p{1.25in}cc}
\tablewidth{0pt}
\tablecaption{
\label{tbl:hizsfd}
Surface Densities of Galaxies of Redshift $z > 5$
}
\tablehead{
\colhead{} &
\colhead{$N$} &
\colhead{$\sigma(N)$}
\\
\colhead{$AB(16,000)$} &
\colhead{(arcmin$^{-2}$)} &
\colhead{(arcmin$^{-2}$)}
}
\startdata
26.0 \dotfill &  0.95 & 0.95 \\
27.0 \dotfill &  2.87 & 1.90 \\
28.0 \dotfill & 18.32 & 3.81 \\
\enddata
\end{deluxetable}

\clearpage

\begin{deluxetable}{p{2.00in}cc}
\tablewidth{0pt}
\tablecaption{
\label{tbl:hizsfd2}
Surface Densities of Galaxies of Redshift $z > 10$
}
\tablehead{
\colhead{} &
\colhead{$N$} &
\colhead{$\sigma(N)$}
\\
\colhead{Magnitude} &
\colhead{(arcmin$^{-2}$)} &
\colhead{(arcmin$^{-2}$)}
}
\startdata
$AB(16,000)=28.0$ \dotfill &  3.58 & $^{+1.90}_{-0.95}$ \\
$AB(22,200)=24.0$ \dotfill &  1.01 & $^{+0.90}_{-0.90}$ \\
\enddata
\end{deluxetable}

\clearpage

\begin{deluxetable}{p{2.0in}cccc}
\small
\tablewidth{0pt}
\tablecaption{
\label{tbl:hie}
Early-Type Galaxies of Redshift $z > 1$
}
\tablehead{
\colhead{Object} &
\colhead{$\alpha$} &
\colhead{$\delta$} &
\colhead{$AB(16,000)$} &
\colhead{$z_{\rm phot}$}
}
\startdata
SB--NI--0608--0372 \dotfill & 22:32:51.09 & $-$60:39:09.75 & 21.66 & 1.35 \\
SB--NI--0648--0899 \dotfill & 22:32:50.69 & $-$60:38:30.26 & 23.19 & 1.99 \\
SB--NI--0886--0718 \dotfill & 22:32:48.26 & $-$60:38:43.78 & 25.85 & 3.24 \\
SB--NI--0471--0941 \dotfill & 22:32:52.49 & $-$60:38:27.08 & 27.83 & 5.10 \\
SB--NI--0844--0698 \dotfill & 22:32:48.69 & $-$60:38:45.28 & 27.54 & 7.16 \\
\enddata
\end{deluxetable}

\clearpage

\begin{deluxetable}{p{1.25in}cc}
\tablewidth{0pt}
\tablecaption{
\label{tbl:hiesfd}
Surface densities of Early-Type Galaxies of Redshift $z > 1$
}
\tablehead{
\colhead{} &
\colhead{$N$} &
\colhead{$\sigma(N)$}
\\
\colhead{$AB(16,000)$} &
\colhead{(arcmin$^{-2}$)} &
\colhead{(arcmin$^{-2}$)}
}
\startdata
25.0 \dotfill & 1.90 & $^{+1.90}_{-0.95}$ \\
26.0 \dotfill & 2.85 & $^{+1.90}_{-0.95}$ \\
27.0 \dotfill & 2.85 & $^{+1.90}_{-0.95}$ \\
28.0 \dotfill & 5.33 & $^{+1.90}_{-2.85}$ \\
\enddata
\end{deluxetable}

\clearpage

\begin{deluxetable}{p{1.5in}ccccp{1.05in}ccc}
\small
\tablewidth{0pt}
\tablecaption{
\label{tbl:comp}
Comparison of Photometric Redshift Measurements
}
\tablehead{
\multicolumn{4}{c}{Current Analysis} &
\colhead{} &
\multicolumn{4}{c}{Analysis of \citet{Benitez99}}
\\
\cline{1-4} \cline{6-9}
\\
\colhead{Object} &
\colhead{$\!\!\!\!\!AB(16,000)$} &
\colhead{$z_{\rm phot}$} & 
\colhead{Type} &
\colhead{} &
\colhead{Object} &
\colhead{$\!\!\!\!\!AB(16,000)$} &
\colhead{$z_{\rm phot}$} &
\colhead{Type}
}
\startdata
SB--NI--0181--0870 \dotfill & 21.20 & 0.98 & E/S0 & & NIC3--ET1 \dotfill & 21.24 & 1.41 & E   \\
SB--NI--0440--0353 \dotfill & 24.34 & 2.23 & Sbc  & & NIC3--ET2 \dotfill & 24.50 & 1.55 & E   \\
SB--NI--0608--0372 \dotfill & 21.66 & 1.35 & E/S0 & & NIC3--ET3 \dotfill & 21.79 & 1.66 & E   \\
SB--NI--0648--0899 \dotfill & 23.19 & 1.99 & E/S0 & & NIC3--ET4 \dotfill & 23.18 & 1.94 & E   \\
SB--NI--0596--1002 \dotfill & 23.45 & 1.93 & Sbc  & & NIC3--SP1 \dotfill & 23.53 & 2.1  & Sab \\
SB--NI--0837--1023 \dotfill & 22.10 & 1.39 & Scd  & & NIC3--SP2 \dotfill & 22.33 & 1.2  & Sbc \\
\enddata
\end{deluxetable}

\end{document}